\documentstyle[12pt]{article}
\begin{document}
\begin{titlepage}
\begin{center}
{\Large \bf  Renormalizations in Softly Broken SUSY \protect\\[0.5cm]
Gauge Theories}

%\vspace{5mm}
\vglue 10mm
{\bf L.V.~Avdeev\footnote{E-mail: avdeevL@thsun1.jinr.dubna.su},~
 D.I.~Kazakov\footnote{E-mail: kazakovd@thsun1.jinr.dubna.su}~
 and~ I.N.~Kondrashuk\footnote{E-mail: ikond@thsun1.jinr.dubna.su}}

%\vspace{5mm}
\vglue 5mm
{\it Bogoliubov Laboratory of Theoretical Physics, \\
Joint Institute for Nuclear Research, \\
141 980 Dubna, Moscow Region, RUSSIA }
\end{center}

%\vspace{5mm}
\vglue 20mm
\begin{abstract}
The supergraph technique for calculations in
supersymmetric gauge theories where supersymmetry is broken in a "soft" way
(without introducing quadratic divergencies) is reviewed.
By introducing an external spurion field the  set of Feynman rules is
formulated and explicit connections between the UV counterterms of a
softly broken and rigid SUSY theories are found.
It is shown that the renormalization constants of softly broken SUSY gauge
theory  also become external superfields depending on the spurion field.
Their explicit form repeats that of the constants of a rigid theory with the
 redefinition of the couplings.  The method  allows us to reproduce all known
 results on the renormalization of  soft couplings and masses in a softly
broken theory.  As an example  the renormalization group functions  for
soft couplings and masses in the Minimal Supersymmetric Standard Model up to
the three-loop level are calculated.
\end{abstract}

%\vspace{5mm}
\vglue 10mm
\noindent PACS numbers:
11.10Gh, 11.10Hi,11.30Pb

\end{titlepage}

\section{Introduction}

Supersymmetry, if it exists, must be broken.
The gauge theory  with softly broken supersymmetry has been widely
studied.  To break supersymmetry without destroying the
renormalization properties of SUSY theories, in particular the
non-renormalization theorems and the cancellation of quadratic divergencies,
one has to introduce the so-called soft terms~\cite{spurion}. They
include the bilinear and trilinear  scalar couplings and the mass terms
for scalars and gauginos.

A powerful method for  studying SUSY theories which keeps supersymmetry
manifest is the supergraph technique~\cite{supergraph,supergraph2}. It is
also applicable to softly broken SUSY models by using the "spurion" external
superfields~\cite{spurion,spurion2,Scholl}. As has been shown by
Yamada~\cite{Yamada} with the help of the spurion method the calculation of
the $\beta$ functions of soft SUSY-breaking terms is a much simpler task
than in the component approach. Using the spurion technique he has
derived very efficient rules which allow him to calculate the soft
term divergencies starting from those of a rigid theory in low orders
of perturbation theory.

However, these rules need  some modification since, as mentioned by
Yamada~\cite{Yamada}, there might appear some problems in the calculation of
vector vertices because they contain chiral derivatives acting on external
lines.

In the present paper we develop  such a modification of the spurion technique
in gauge theories which takes care of the above mentioned problem. We
 formulate the Feynman rules and  show that the ultraviolet divergent parts
of the Green functions of a softly broken SUSY gauge theory possess  the
factorization property  being proportional to those  of a rigid theory.

In general, we consider a softly broken supersymmetric gauge theory
as a rigid SUSY theory imbedded into the external space-time independent
superfield, so that all the parameters as the couplings and masses
become external superfields.
This approach to a softly broken sypersymmetric theory allows us to use
remarkable mathematical properties of $N=1$ SUSY theories such as
non-renormalization theorems, etc.
We show that the renormalization procedure in a softly broken SUSY
gauge theory can be performed in exactly the same way as in a rigid theory
with the renormalization constants being external superfields. They are
related to the corresponding  renormalization constants of a rigid theory by
the coupling constants redefinition. This  allows us to find explicit
relations between the  renormalizations  of soft  and rigid  couplings.

Throughout the paper we assume the existence of some gauge and SUSY invariant
regularization. Though it is   some problem by itself, we do not consider it
here.  Provided the rigid theory is well defined, we consider the
modifications which appear due to the presence of  soft SUSY breaking terms.
To be more precise, in the following sections when discussing one, two and
three loop calculations of the renormalization constants we have in mind
dimensional reduction and the minimal subtraction scheme. Though dimensional
reduction is not self-consistent in general, it is safe to use it in low
orders and all the actual calculations are performed in the framework of
dimensional reduction~\cite{Vaughn,Jones,Jack,Ferreira}. Nevertheless,
our main formulae have  general validity provided the invariant
procedure exists.

\section{Softly Broken Pure  $N=1$ SUSY  Yang--Mills \protect\\ Theory }

Let us first consider  pure $N=1$ SUSY Yang-Mills theory with a simple
gauge group. The Lagrangian  of a rigid theory is given by
\begin{eqnarray}
{\cal L}_{rigid} &=& \int d^2\theta~\frac{1}{4g^2}{\rm Tr}W^{\alpha}W_{\alpha}
+ \int d^2\bar{\theta}~\frac{1}{4g^2}{\rm Tr}\bar{W}^{\alpha}\bar{W}_{\alpha}.
 \label{rigidlag}
\end{eqnarray}
Here  $W^{\alpha}$ is the field strength chiral superfield defined by
\begin{eqnarray}
W^{\alpha} = \overline{D}^2 \left(
e^{-V} D^{\alpha}e^V \right), \nonumber
\end{eqnarray}
where $V^a_b = V^A \left(T^A\right)^a_b$ and $T^A$
are the generators  of the gauge group $G.$

To  perform a soft SUSY breaking, one can introduce a gaugino mass
term. This is the only term allowed in a pure gauge theory which does
not break a gauge invariance.  The soft SUSY breaking term is
$$ -{\cal L}_{soft-breaking} =
\frac{m_A}{2}\lambda\lambda + \frac{m_A}{2}\bar \lambda \bar \lambda,
$$
where $\lambda$ is  the gaugino field. To rewrite it in terms of
superfields, let us  introduce an external spurion superfield
 $\eta=\theta^2$, where $\theta$ is a grassmannian parameter. The
softly broken Lagrangian  can now be written as
\begin{eqnarray}
 {\cal L}_{soft} &=& \int d^2\theta~\frac{1}{4g^2}(1-2\mu\theta^2) {\rm
Tr}W^{\alpha}W_{\alpha} + \int
 d^2\bar{\theta}~\frac{1}{4g^2}(1-2\bar{\mu}\bar{\theta}^2) {\rm
Tr}\bar{W}^{\alpha}\bar{W}_{\alpha}.  \label{soft}
\end{eqnarray}
In terms of component fields the interaction with external spurion
 superfield leads to a gaugino mass equal to $m_A=\mu$, while the gauge
field remains massless.  This external chiral superfield can be
considered as a vacuum expectation value of a dilaton superfield
emerging from supergravity, however, this is not relevant to further
consideration.

\subsection{The Feynman rules}

Consider now the Feynman rules corresponding to the Lagrangian
(\ref{soft}). For this purpose it is useful   to rewrite the integral
over the chiral superspace in (\ref{soft}) in terms of an integral over
the whole superspace. One has
\begin{eqnarray}
{\cal L}_{soft} &=& \int d^2\theta d^2\bar{\theta}
\left(\frac{1}{4g^2}(1-2\mu\theta^2) {\rm Tr}e^{-V}D^\alpha
e^V\bar{D}^2e^{-V}D_\alpha e^V  \right.\nonumber \\
&&\left. +  \frac{1}{4g^2}(1-2\bar{\mu}\bar{\theta}^2) {\rm
Tr}e^{-V}\bar{D}^\alpha e^V D^2 e^{-V}\bar{D}_\alpha e^V \right).
\label{full}
\end{eqnarray}

Consider the quadratic part of the action
\begin{eqnarray*}
S^{(2)} &=& \int d^4xd^2\theta~d^2\bar{\theta}{\rm Tr}~
\frac{1}{4g^2}\left(1 - 2\mu\theta^2 \right)
D^{\alpha}V(x,\theta,\bar{\theta})\bar{D}^2D_{\alpha}V(x,\theta,\bar{\theta}) \nonumber\\
&+& \int d^4xd^2\theta~d^2\bar{\theta}{\rm Tr}~
\frac{1}{4g^2}\left(1 - 2\mu\bar{\theta}^2 \right)
\bar{D}^{\dot{\alpha}}V(x,\theta,\bar{\theta})D^2
D_{\dot{\alpha}}V(x,\theta,\bar{\theta}).  \label{quadr}
\end{eqnarray*}

To find the propagator, one first has to fix the gauge. In a rigid
theory the gauge fixing condition is usually taken as
\begin{eqnarray}
\bar{D}^2V=f, ~~~ D^2V = \bar{f}. \label{14}
\end{eqnarray}
This is equivalent to adding to the action (\ref{rigidlag}) a
gauge-noninvariant term proportional to
\begin{equation}
{\cal L}_{gauge-fixing}={\rm Tr}~\frac{1}{\alpha}\bar{f}f =
{\rm Tr}~\frac{1}{2\alpha}\left(\bar{f}f + f\bar{f}\right).
\end{equation}

By joining this together the quadratic part of the rigid action takes
the form
$$
S^{(2)}_{rigid}= \int d^2\theta~d^2\bar{\theta}\left[ \frac{1}{4g^2}{\rm
Tr}D^{\alpha}V\bar{D}^2D_{\alpha}V + \frac{1}{4g^2}{\rm Tr}\bar
D^{\alpha}VD^2\bar D_{\alpha}V - \frac{1}{2\alpha}D^2V\bar{D}^2V
\right].
$$
It can also be rewritten as
\begin{eqnarray*}
\frac{4}{g^2}
\int d^2\theta~d^2\bar{\theta}{\rm Tr}V\left[\Pi_{1/2} +
\frac{1}{\alpha}\Pi_0\right]{\partial}^2V,
\label{Qprwp}
\end{eqnarray*}
where we have introduced the projectors~\cite{West}
\begin{eqnarray*}
\Pi_{1/2} = - \frac{D^{\alpha}\bar{D}^2D_{\alpha}}{8{\partial }^2},~~~
\Pi_0 = \frac{D^2\bar{D}^2 + \bar{D}^2D^2}{16{\partial}^2}, \nonumber
\end{eqnarray*}

The inverse operator then gives the vector propagator
\begin{eqnarray*}
\left<V^a(z_1)V^b(z_2)\right>_{rigid} = \frac{g^2}{4}\frac{\Pi_{1/2} +
\alpha\Pi_0}{{\partial}^2}\delta^{(8)}(z_1 - z_2)\delta^{ab},
\label{Pvr}
\end{eqnarray*}

The next task related to the determination of the vector propagator
is to find the associated ghost term.  Under the gauge
transformation the  superfield $V$ is transformed as
\begin{eqnarray*}
e^V  \rightarrow e^{-i\Lambda^{\dagger}}e^Ve^{i\Lambda}, \nonumber
\end{eqnarray*}
or in the infinitesimal form
\begin{eqnarray*}
&& V \rightarrow V + H(V)\Lambda + \bar{H}(V)\bar{\Lambda} \nonumber\\
&& = V +  \left(-\frac{V}{2}\right) \bigwedge
\left[\left(\Lambda+\overline{\Lambda}\right) + \coth\left(\frac{V}{2}
\right) \bigwedge \left(\Lambda-\overline{\Lambda}\right)\right]. \label{15}
\end{eqnarray*}
Consequently, the associated ghost term is
\begin{eqnarray}
{\cal L}_{ghost}&=&
\int d^2\theta~{\rm Tr}~\frac{1}{g^2}b\bar{D}^2\left(H(V)c +
\bar{H}(V)\bar{c}\right) + \int d^2\bar{\theta}~{\rm
Tr}~\frac{1}{g^2}\bar{b}D^2\left(H(V)c + \bar{H}(V)\bar{c}\right)
\nonumber\\ &=&
\int d^2\theta d^2\bar{\theta}~{\rm Tr}~\frac{1}{g^2} \left(b+
\overline{b}\right) \left(-
\frac{V}{2}\right) \bigwedge \left[\left(c+\overline{c}\right) +
\coth\left(\frac{V}{2} \right) \bigwedge
\left(c-\overline{c}\right)\right], \label{16}
\end{eqnarray}
where $b$ and $c$ are the Faddeev--Popov ghost chiral superfields.

For a system in the  external spurion field we can change the gauge
fixing condition (\ref{14}) in order to take into account the
interaction of the associated ghost fields with the spurion. The reason
for this is the appearance of these terms in due course of
renormalization. Here, we depart from the Feynman rules of
ref.\cite{Yamada}. Our aim is also to get the same common spurion
factor for the ghost propagator as for the vector one, as it will be
clear below.  For these purposes we choose the gauge-fixing condition
in the form
\begin{eqnarray*}
 &&\bar{D}^2\frac{1}{g^2}\left(1 -
\mu\theta^2 - \mu\bar{\theta}^2\right)V=f, ~~~ D^2\frac{1}{g^2}\left(1
- \mu\theta^2 - \mu\bar{\theta}^2\right)V = \bar{f}.\nonumber
\end{eqnarray*}
or
\begin{eqnarray}
\bar{D}^2\frac{1}{{\tilde g}^2}V=f, ~~~ D^2\frac{1}{{\tilde g}^2}V = \bar{f}.
\label{17}
\end{eqnarray}
where we have introduced the notation
\begin{eqnarray}
\tilde{g}^2 = g^2\left(1 + \mu\theta^2 + \mu\bar{\theta}^2 +
2\mu^2\theta^2 \bar{\theta}^2 \right).  \label{Tildeg}
\end{eqnarray}
Note that due to the grassmannian origin of $\theta$
\begin{eqnarray*}
\left(1 - \mu\theta^2 - \mu\bar{\theta}^2\right)
\left(1 + \mu\theta^2 + \mu\bar{\theta^2} + 2\mu^2\theta^2
\bar{\theta^2} \right) = 1. \nonumber
\end{eqnarray*}

Now the gauge-fixing term can be written in the standard form
\begin{eqnarray}
{\cal L}_{gauge-fixing}&=& -\int d^2\theta d^2\bar{\theta}{\rm
Tr}~\frac{1}{\xi}\bar{f}f,
\end{eqnarray}
where $\xi$  might be a  real constant   superfield. In what follows we take
\begin{eqnarray*}
 \xi = \frac{\alpha}{{\tilde g}^2}, \nonumber
\end{eqnarray*}
which leads to  the following quadratic part of the action in a softly
broken theory
\begin{eqnarray}
S^{(2)}_{soft}&=&\int d^4xd^2\theta~d^2\bar{\theta}{\rm
Tr}\left[ \frac{1}{4g^2}\left(1 - 2\mu\theta^2 \right)
D^{\alpha}V\bar{D}^2D_{\alpha}V \right. \label{Qpsw} \\
&&+ \left.
 \frac{1}{4g^2}\left(1 - 2\mu\bar{\theta}^2 \right)
\bar{D}^{\dot{\alpha}}VD^2\bar{D}_{\dot{\alpha}}V +
\frac{\tilde{g}^2}{2\alpha}
\left(\bar{D}^2\frac{1}{\tilde{g}^2}V\right)
\left(D^2\frac{1}{\tilde{g}^2} V\right)
\right]. \nonumber
\end{eqnarray}

One now has to find the inverse operator. It can be done analogously
to the rigid case by introducing the projection operators. One should
take into account, however, that while integrating the covariant
derivatives by part they may also act on the spurion field. Strictly
speaking, this leads to additional terms in the propagator which are,
however, suppressed by powers of momenta and are inessential in the
analysis of UV divergences. To understand it better, one can  treat the
spurion terms in eq.(\ref{Qpsw}) as interactions, and to find the
propagator, one has to take into account an infinite chain of this kind
of insertions adding up into a geometrical progression. It
can be done by the method of Ref.~\cite{Scholl}. If in due course of
this procedure some of the covariant derivatives act on a spurion
field, one does not get enough powers of momenta to cancel the
denominator and the resulting terms are thus suppressed for a high
momentum square. Thus, as it has been also noted in Ref.~\cite{Yamada},
for the calculation of the divergent parts of the Green functions
it is sufficient to take the part of the vector propagator
where the chiral derivatives do not act on spurions.
%the soft theory the essential part of the vector propagator arises when

Having this in mind, one can proceed in full analogy with the rigid
theory, replacing the coupling $g^2$ by ${\tilde g}^2$, and get
\begin{eqnarray*}
\left<V^a(z_1)V^b(z_2)\right>_{soft} =
\frac{\tilde{g}^2}{4}\frac{\Pi_{1/2} +
\alpha\Pi_0}{{\partial}^2}\delta^{(8)}(z_1 - z_2)\delta^{ab} +
{\rm irrelevant~~terms}, \label{Pvs}
\end{eqnarray*}
where by irrelevant terms we mean the ones decreasing faster than
$1/p^2$ for large $p^2$.
Thus, we get a simple relation between the vector propagators of  soft
and rigid theories:
\begin{eqnarray}
&&\left<V(x_1,\theta_1,\bar{\theta_1})V(x_2,\theta_2,\bar{\theta_2})
\right>_{soft}= \\
 &&= \frac{{\tilde g}^2}{g^2} \left<V(x_1,\theta_1,
\bar{\theta_1})V(x_2,\theta_2,\bar{\theta_2})\right>_{rigid}
+ {\rm irrelevant~~terms}, \nonumber
\end{eqnarray}
 Notice that contrary to ref.~\cite{Yamada} this relation is valid for any
choice of $\alpha$.

The change of the gauge fixing condition leads to the change of the
ghost Lagrangian (\ref{16}). It becomes
\begin{eqnarray}
{\cal L}_{ghost} &=&\int d^2\theta~{\rm Tr}~b\bar{D}^2
\left(\frac{1}{\tilde{g}^2} \left(H(V)c +
\bar{H}(V)\bar{c}\right)\right) \nonumber \\
&& \hspace*{3cm} + \int
d^2\bar{\theta}~{\rm Tr}~\bar{b}D^2\left(\frac{1}{\tilde{g}^2}
\left(H(V)c + \bar{H}(V)\bar{c}\right)\right)
\nonumber\\
&=&\int d^2\theta d^2\bar{\theta}~{\rm Tr}~\frac{1}{\tilde{g}^2}
\left(b+\overline{b}\right) \left(- \frac{V}{2} \bigwedge
\Bigl[\left(c+\overline{c}\right) +
\coth\left(\frac{V}{2} \right) \bigwedge
\left(c-\overline{c}\right)\Bigr]\right). \label{18}
\end{eqnarray}

As one can see from  eq. (\ref{18}), the situation with the ghost
propagator and the vertices that contain ghost superfields is more simple.
This is due to the absence of chiral derivatives in the ghost-vector
interaction. Nevertheless, when calculating the inverse operator
one has to repeat the same procedure with the covariant derivatives
arising from the source terms. In complete analogy with the  case
of the vector propagator the essential part of the ghost propagator
is obtained by assuming that the covariant derivatives do not act on the
spurion field. The other terms are suppressed by the powers of momenta.
Hence, one again has
\begin{equation}
\left<G(z_1)\bar{G}(z_2)\right>_{soft} =
\left(\tilde{g}^2/g^2\right) \left<G(z_1)\bar{G}(z_2)\right>_{rigid}
 +~~{\rm irrelevant~~terms}, \label{Gps}
\end{equation}
where $G$ stands for any ghost superfield.

Thus, to perform the analysis of the divergent part of the diagrams in
a soft theory, one has to use the same propagators as in a rigid theory
multiplied by the factor ${\tilde g}^2/g^2$. It is also obviously true
for any vertex of the ghost-vector interactions of the softly broken
theory (\ref{18}). Each vertex of this type has to be
multiplied by the inverse factor $g^2/{\tilde g}^2$.

The situation is less obvious with the vector vertices. Here, one has
two terms as in eq.(\ref{soft}) which differ only in the order
of covariant derivatives, one being complex conjugated to the other.
Performing grassmannian algebra in any diagram, one can always replace
one term by the other; so actually one has only one  term. This is what
we have in a rigid theory. In a soft theory the situation is similar
but the first term is multiplied by a factor $\left(1 -
2\mu\theta^2\right),$ while the second by the
complex conjugated one $\left(1 -  2\mu\bar{\theta}^2\right)$.
Hence, one can consider the vector vertices of only one type, as in
a rigid theory, multiplied by a factor $\left(1 - \mu\theta^2 -
\mu\bar{\theta}^2\right)$, i.e. by $g^2/{\tilde g}^2$ as in the case
of the ghost-vector vertices.

Thus, we see that any element of the Feynman rules for a softly broken
theory  coincides with the corresponding element of a rigid theory
multiplied by a common factor which is a polynomial in the grassmann
coordinates.

\subsection{The ultraviolet counterterms}

Consider now the Green functions of a softly broken theory constructed with
the help of the Feynman rules derived in the previous section.  Due to the
presence of a factor $g^2/{\tilde g}^2$ in the
vertices and of the inverse one in the propagators, any Feynman diagram
in a broken theory is given by that of a rigid theory with the spurion
factors on the lines and vertices. Our aim is to show that these
polynomials in grassmann variables can freely flow through the diagram
and are finally collected in front of a  diagram as a common factor.

Indeed, consider the corresponding grassmannian integration. The usual
procedure of evaluation of grassmannian integrals in a rigid theory is
the following:
\begin{enumerate}
\item[i)] If there are no chiral derivatives $D_\alpha$ or
$\bar{D}_\alpha$ on the line, it is proportional to the
$\delta$-function and can be shrunk to a point.
\item[ii)] If there is some number of derivatives, one  can remove them
from a line integrating by parts.
\item[iii)] Repeating this procedure  several  times and using the
properties of chiral derivatives
$$\{D_\alpha,  D_\beta\}= \{\bar{D}_\alpha,\bar{D}_\beta\}=0,  \ \ \ \
D_\alpha D_\beta = \epsilon_{\alpha \beta}D^2, \ \ \ \ D^2{\bar
D}^2D^2=16p^2D^2, \ \ {\rm etc.} $$
to reduce their number, one can finally come to a final integration
where all the chiral derivatives are concentrated on one line. The
powers of momenta which appear in due course of this procedure cancel
some propagator lines so that the resulting diagram diverges.
\item[iv)] The final integration is performed with the help of the
relation
\begin{equation}
\delta_{12}\bar{D}^2D^2\delta_{12}=16\delta_{12}.
\label{rule}
\end{equation}
\end{enumerate}

Consider what happens if one adds grassmannian polynomials on the lines and
vertices.  Suppose first that all chiral derivatives inside a graph are
contracted and integrated inside the graph so that there are no external
derivatives left. If one of these internal
derivatives acts on a spurion field it  creates  additional
$\theta^\alpha$ instead of $D^\alpha$. Then either this $\theta^\alpha$
survives till the last integration, or some other $D_\alpha$  acts on
it.  Thus, at the last step of integration  one has instead of
(\ref{rule})  either
$$
\delta_{12}\theta^{\alpha}_2\bar{D}^2_1D^\alpha_1\delta_{12} \ \ \ {\rm or}
 \ \ \ \delta_{12}\bar{D}^2_1\delta_{12}.
$$
It is easy to see  that both the expressions are zero. Thus, the
configuration when the chiral derivative acts on a spurion field gives
no contribution.  This means that the spurion field flows through the
diagram freely and can be factorized.

Consider now the case when a graph contains some derivatives that act on
external lines when  performing the grassmannian integration. This case
corresponds to vector interactions.

Let some covariant derivative,
which originally was inside the diagram, pass through the diagram
during the  above mentioned procedure and act on the external vector
line.  We can trace this derivative in the diagram of the soft theory.
On its way through the diagram it can meet some "spurion" factor in
a vertex or a propagator. Integrating by parts one has two terms.
The first one, when the chiral derivative acts on a spurion, and the
second one, when it acts on the neighbouring propagator.
When the covariant derivative does not act on a spurion it goes
through the diagram like in a rigid theory and acts on external line.
As a result one gets the expression
$\tilde{g}^2 \left(D_{\alpha}V^m\right)$. In the other case, when the
covariant derivative acts on the spurion,
the number of covariant derivatives is reduced compared to
the rigid theory. Then, either one does not have enough of them to
cancel the powers of momenta in the denominator or to get the non-zero
answer or one has the same diagram as in the rigid case, but with the
factor $D{\tilde g}^2$, that goes outside the diagram.  Independently
of the position of this factor inside the diagram, the final expression
one gets after the calculation of the diagram contains
$\left(D_{\alpha}\tilde{g}^2\right)V^m$ which together with the usual
contribution, gives $D_{\alpha}\left(\tilde{g}^2V^m\right)$ instead of
$g^2D_{\alpha}V^m$ in the rigid case.

All together these factors $\tilde{g}^2$, which are met by the
chiral derivative on its way to the external line of the
diagram,  are collected in a monom $R(\tilde{g}^2).$  Therefore, the
final expression for the diagram contains
$D_{\alpha}\left(R(\tilde{g}^2)V^m\right)$.
The same conclusion is true for all four derivatives of the
divergent vector diagram that act on the external lines. Thus,
the counterterm to this diagram of the soft theory is
\begin{eqnarray}
R_1(\tilde{g}^2)V^mD^{\alpha}\left( R_2(\tilde{g}^2)V^n\right)
\bar{D}^2\left(R_3(\tilde{g}^2)V^kD_{\alpha} R_4(\tilde{g}^2)V^l\right),
\label{C}
\end{eqnarray}
where $R_1$, $R_2$, $R_3$ and $R_4$ are some monoms in
$\tilde{g}^2.$ This expression should be compared with
the counterterm for the corresponding diagram of the rigid theory
\begin{eqnarray}
R(g^2)V^mD^{\alpha}V^n\bar{D}^2V^kD_{\alpha}V^l,
\nonumber
\end{eqnarray}
where
\begin{eqnarray}
R_1(g^2)R_2(g^2)R_3(g^2)R_4(g^2)  = R(g^2).  \nonumber
\end{eqnarray}
The momentum integral for the diagram in a soft
theory is identical to that in a rigid one since the algebraic
operations  with  the covariant derivatives coincide
for the both cases.

We now want to argue that the monoms $R_i(\tilde{g}^2)$ can stand
anywhere in eq.(\ref{C}), i.e. that the spurions can flow
through the diagram. Indeed, in a rigid theory the counterterms are
absorbed into the redefinition of the field $V$ and
the coupling $g^2$
\begin{eqnarray} V_B = Z^{1/2}_V(g^2)V, \nonumber\\
g_B^2 = Z_g(g^2)g^2 \nonumber
\end{eqnarray}
In a soft theory $R_i$ are the functions of the external
spurion superfield and the renormalization constants become
these functions too.
Collecting all counterterms of the type (\ref{C}), one should get
an expression which being written in terms of the bare quantities
obeys the gauge invariance, namely
\begin{eqnarray}
{\rm
Tr}~R_1'(\tilde{g}^2)e^{-V_B}D^{\alpha}R_2'(\tilde{g}^2)
e^{V_B}\bar{D}^2\left(R_3'(\tilde{g}^2)
e^{-V_B}D_{\alpha}R_4'(\tilde{g}^2)e^{V_B}\right),
\label{Ba}
\end{eqnarray}
where the primed monoms mean that partly they are absorbed into the
renormalization of the field $V_B$.

Now it is clear that the covariant derivatives in eq.(\ref{Ba}) do not
act on monoms $R_i({\tilde g}^2)$. Indeed, if one of the covariant
derivatives $D_\alpha $ in eq.(\ref{Ba}) acts on the momons $R_2'$
or $R_4'$, the two exponents cancel each other and one is left
with ${\rm Tr}~W^{\alpha}$ that is equal to zero. In the case when
the both chiral derivatives in eq. (\ref{Ba}) act on $R_2'$ and $R_4'$
in both cases the exponents cancel and one has a divergent constant
that should be absorbed into the renormalization of the vacuum density.

Hence, one can take the monoms $R_2'$ and $R_4'$ out of the covariant
derivatives $D_\alpha $. The other derivative, $\bar{D}^2$, does not
act on the monom for the reason that
$e^{-\tilde{V}_B}D^{\alpha}e^{\tilde{V}_B}\bar{\theta}^2 =0$
so that in the bracket in eq.(\ref{Ba}) only the chiral spurion
$\eta=\theta^2$ survives and $\bar{D}^2$ does not act on it.

Thus, we come to a conclusion that the monoms can be factorised in front
of the expression in eq.(\ref{Ba}) just like in the rigid theory
and the only difference between the rigid and softly broken
theories is that the coupling constant  $g^2$ should be replaced
by ${\tilde g}^2$, i.e.
\begin{eqnarray}
\tilde{Z}_i = Z_i\left(g^2 \rightarrow \tilde{g}^2\right), \nonumber
\end{eqnarray}

\section{Softly Broken SUSY Gauge Theory with Chiral Matter}

Consider now a rigid SUSY gauge theory with chiral matter.  The Lagrangian
written in terms of superfields looks like
\begin{equation}
{\cal L}_{rigid}  = \int d^2\theta d^2\bar{\theta} ~~\bar{\Phi}^i
(e^{V})^j_i\Phi_j + \int d^2\theta ~~{\cal W} + \int d^2\bar{\theta}
~~\bar{\cal W},
\end{equation}
where the superpotential ${\cal W}$ in a  general form is\footnote{ To avoid
further complications (see, e.g.  \cite{Yamada}) we do not introduce
linear terms in eqs.(\ref{rigid},\ref{sofl}). This corresponds to the case
when none of the fields goes into vacuum. In the absence of singlets this is
guaranteed by the gauge invariance imposing some restrictions on the
couplings.}
\begin{equation} {\cal  W}=\frac{1}{6}\lambda^{ijk}\Phi_i\Phi_j\Phi_k
+\frac{1}{2} M^{ij}\Phi_i\Phi_j.\label{rigid}
\end{equation}
The SUSY  breaking terms which satisfy the requirement of "softness"  can be
written as
\begin{equation} -{\cal L}_{soft-breaking}=\left[\frac 16 A^{ijk}
\phi_i\phi_j\phi_k+ \frac 12 B^{ij}\phi_i\phi_j +h.c.\right]
+(m^2)^i_j\phi^{*}_i\phi^j,\label{sofl}
\end{equation}

Like in the case of a pure gauge theory the soft terms (\ref{sofl}) can be
written down in terms of superfields by using the external spurion
field.  Thus, the full  Lagrangian for the softly broken theory can be
written as
\begin{eqnarray}
{\cal L}_{soft}&=&\int d^2\theta
d^2\bar{\theta} ~~\bar{\Phi}^i(\delta^k_i -(m^2)^k_i\eta
\bar{\eta})(e^V)^j_k\Phi_j   \label{ssofl2} \\ &+& \int  d^2\theta
\left[\frac 16 (\lambda^{ijk}-A^{ijk} \eta)\Phi_i\Phi_j\Phi_k+ \frac 12
(M^{ij}-B^{ij}\eta ) \Phi_i\Phi_j \right] +h.c.  \nonumber
\end{eqnarray}

\subsection{The Feynman rules}

The Lagrangian (\ref{ssofl2}) allows one to write down the
Feynman rules for the matter field propagators and vertices in a soft
theory.  We start with the propagator of the chiral field. For the
purpose of the analysis of divergences one can ignore the mass terms
$M^{ij}$ and $B^{ij}$, since in the minimal scheme they do not
contribute to the UV divergences.  Then, the quadratic part looks like
\begin{eqnarray}
{\cal L}^{(2)}_{soft}&=&\int d^2\theta d^2\bar{\theta}
~~\bar{\Phi}^i(\delta^j_i -(m^2)^j_i\eta \bar{\eta})\Phi_j.
\nonumber
\end{eqnarray}
The inverse operator is easy to obtain due to the nilpotent character
of the spurion fields. One easily gets
\begin{eqnarray}
\left<\Phi(z_1)_i\bar{\Phi}(z_2)^j\right>_{soft}& =&
(\delta^k_i +\frac 12(m^2)^k_i\eta \bar{\eta})
\left<\Phi(z_1)_k\bar{\Phi}(z_2)^l\right>_{rigid}
(\delta^j_l +\frac 12(m^2)^j_l\eta \bar{\eta}) \nonumber \\
&& +~~{\rm irrelevant~~terms},  \label{prphi}
\end{eqnarray}
where the irrelevant terms again arise from covariant derivatives acting
 on spurions and are suppressed by powers of momenta.

The vector-matter vertices,  according to eq.(\ref{ssofl2}), gain
the factor $(\delta^j_i -(m^2)^j_i\eta \bar{\eta})$ so that if in a
diagram one has an equal number of chiral propagators and
vector-matter vertices the spurion factors cancel.

The chiral vertices of a soft theory, as it follows from
eq.(\ref{ssofl2}), are the same as in a rigid theory with the
Yukawa couplings being replaced by
$$
\lambda^{ijk} \to \lambda^{ijk}-A^{ijk}\eta, \ \ \ \bar \lambda_{ijk}
\to \bar \lambda_{ijk} - \bar A_{ijk} \bar{\eta}.
$$

\subsection {The ultraviolet counterterms}

The structure of the UV counterterms in chiral vertices is similar to
that of the vector vertices, but is simpler due to the absence of
the covariant derivatives on external lines. This corresponds to the
first case considered in the previous section. To get the UV divergent
diagram, the covariant derivatives should not act  on spurion fields,
which means that spurions factorize.
In the diagrams, when the chiral vertices are contracted
with the chiral propagators, this results in the following
effective change of the Yukawa couplings of a softly broken theory
\begin{eqnarray}
\lambda^{ijk} &\to&\tilde{\lambda }^{ijk}=
\lambda^{ijk}-A^{ijk}\eta +\frac 12
(\lambda^{njk}(m^2)^i_n +\lambda^{ink}(m^2)^j_n+\lambda^{ijn}(m^2)^k_n)\eta
\bar \eta,  \label{l1}\\
\bar \lambda_{ijk} &\to&\tilde{\bar \lambda }_{ijk}=
\bar \lambda_{ijk} - \bar A_{ijk} \bar{\eta}+ \frac 12
(\lambda_{njk}(m^2)_i^n
+\lambda_{ink}(m^2)_j^n+\lambda_{ijn}(m^2)_k^n)\eta \bar \eta  .
\label{l2}
\end{eqnarray}

Thus, the UV counterterms of a softly broken theory are obtained from
those of a rigid one by the substitution
$$g^2 \to {\tilde g}^2,\ \ \  \lambda^{ijk} \to {\tilde\lambda }^{ijk},
\ \ \ \lambda_{ijk} \to {\tilde\lambda }_{ijk}.$$

\section{Renormalization of Soft versus Rigid Theory:
\protect \\ the General Case}

The external field construction described above allows one to write down the
renormalization of soft terms starting from the known renormalization
of a rigid theory without any new diagram calculation. The following
statement is valid:

\vspace{0.5cm}

%\newtheorem{statement}{The Statement}
%\begin{statement}
\noindent{\large \bf The Statement} {\it Let a rigid theory
(\ref{rigidlag},\ref{rigid}) be renormalized via introduction of the
renormalization constants $Z_i$, defined within some minimal subtraction
massless scheme. Then, a softly broken theory (\ref{soft},\ref{ssofl2})
is renormalized via introduction of the renormalization superfields
$\tilde{Z}_i$ which are related to $Z_i$ by the coupling constants
redefinition
\begin{equation} \tilde{Z}_i(g,\lambda ,\bar \lambda ,
\dots)=Z_i(\tilde{g}^2,\tilde{\lambda},\tilde{\bar \lambda}, \dots),
\label{Z} \end{equation} where the redefined couplings are
\begin{eqnarray}
\tilde{g}^2&=&g^2(1+\mu \eta+\bar \mu \bar{\eta}+2\mu\bar \mu \eta
\bar{\eta}),\ \ \  \eta=\theta^2, \ \ \ \bar{\eta}=\bar{\theta}^2,
\label{g}\\
\tilde{\lambda}^{ijk}&=&\lambda^{ijk}-A^{ijk}\eta +\frac 12
(\lambda^{njk}(m^2)^i_n +\lambda^{ink}(m^2)^j_n+\lambda^{ijn}(m^2)^k_n)\eta
\bar \eta,  \label{y1}\\
\tilde{\bar \lambda}_{ijk}&=&\bar \lambda_{ijk} -
\bar A_{ijk} \bar{\eta}+ \frac 12 (\lambda_{njk}(m^2)_i^n
+\lambda_{ink}(m^2)_j^n+\lambda_{ijn}(m^2)_k^n)\eta \bar \eta  . \label{y2}
\end{eqnarray} }
%\end{statement}
This allows us to find explicit relations between the  renormalizations  of
soft and rigid  couplings which  is an explicit realization on the level of
final expressions of the rules $A$ and $B$ of ref.\cite{Yamada}.

The renormalization constants for  vector superfields and  gauge couplings
are general superfields, i.e. they depend on $\eta$ and $\bar{\eta}$, while
those for the chiral matter and ghost fields and the parameters of a
superpotential are chiral superfields, i.e. they depend either on $\eta$ or
$\bar{\eta}$.

From eqs.(\ref{Z}) and (\ref{g}-\ref{y2}) it is possible to write down
an explicit differential operator which has to be applied to the $\beta $
functions of a rigid theory in order to get those for the soft terms.
 We first construct this operator in a general case and compare the resulting
expressions with explicit calculations made up to two loops. Then,
using the formulated algorithm we calculate some three-loop soft term
$\beta $ functions. In the last section,  we consider some particular
models.

To simplify the formulas hereafter we use the following notation:
$$\alpha_i = g^2_i/16\pi^2, \ \ y^{ijk}=\lambda^{ijk}/4\pi, \ \
A^{ijk}=A^{ijk}/4\pi. $$

\subsection{Pure gauge theory}

Consider first the gauge couplings $\alpha_i$. One has
\begin{equation}
\alpha_{i}^{Bare}=Z_{\alpha i}\alpha_i \ \ \Rightarrow \ \
\tilde{\alpha}_{i}^{Bare}=\tilde{Z}_{\alpha i}
\tilde{\alpha}_i,  \label{alpha}
\end{equation}
where $Z_{\alpha i}$ is the product of the wave function and vertex
renormalization constants.

Though $\tilde{\alpha}_i$ and $\tilde{Z}_{\alpha i}$ are general superfields, one has to
consider only their chiral parts since the Lagrangian (\ref{rigidlag})
consists of two terms, a chiral and an antichiral, and in each term
only the proper chirality part contributes. Therefore, we consider the
chiral part of eq.(\ref{alpha})
$$
\alpha_{i}^{Bare}(1+m_{A_i}^{Bare}\eta)=\alpha_i(1+m_{A_i}\eta)Z_{\alpha
i}(\tilde{\alpha})\vert_{\bar \eta=0}.
$$
Expanding over $\eta$  one has
\begin{eqnarray}
\alpha_{i}^{Bare}&=&\alpha_i Z_{\alpha i}(\alpha), \label{ab} \\
m_{A_i}^{Bare}\alpha_{i}^{Bare}&=&m_{A_i}\alpha_i Z_{\alpha i}(\alpha
)+\alpha_i D_1 Z_{\alpha i}, \label{m}
\end{eqnarray}
where the operator  $D_1$ extracts the linear w.r.t. $\eta$ part of
$Z_{\alpha i} (\tilde{\alpha})$. Due to eqs.(\ref{g}-\ref{y2}) the explicit
form of $D_1$ is
$$
D_1=m_{A_i}\alpha_i\frac{\partial}{\partial \alpha_i} \ .  \label{D1}
$$
Combining eqs.(\ref{ab}) and  (\ref{m}) one gets
\begin{equation}
m_{A_i}^{Bare}=m_{A_i}+ D_1 \ln Z_{\alpha i} \ \ \ .  \label{mgaug}
\end{equation}

To find the corresponding $\beta$ functions one has to differentiate
eqs.(\ref{ab}) and (\ref{mgaug}) w.r.t. the scale factor having in mind that
the operator $D_1$ is scale invariant. This gives
\begin{equation}
\beta_{\alpha i}=\alpha_i \gamma_{\alpha i}, \ \ \
 \beta_{m_{A i}}=D_1\gamma_{\alpha i} ,\label{ai}
\end{equation}
where $\gamma_{\alpha i}$ is the logarithmic derivative of $\ln Z_{\alpha
i}$ equal to the anomalous dimension of the vector superfield in some
particular gauges.

This result is in complete correspondence with
that of Ref.~\cite{Shifman}. Indeed, from equations
(\ref{ai}) one can derive that
the ratio $m_A\alpha/\beta_{\alpha}$ is renormalization group
invariant.

\subsection{Chiral matter}

Consider now the chiral matter.  Due to the non-renormalization theorems,
there is no renormalization of a superpotential. This means that the only
renormalization comes from the wave functions.  The corresponding term in the
Lagrangian including renormalizations looks like
\begin{equation}
\int  d^2\theta d^2\bar \theta ~~\bar \Phi^i \tilde{Z}_i^j \Phi_j,
\end{equation}
where the renormalization superfield $\tilde{Z^i_j}$ now  has
a decomposition
\begin{equation}
\tilde{Z}^i_j=\tilde{\bar Z}^{i}_{\phi k}\left( \delta^k_l+\Delta^k_l
\eta \bar \eta \right)\tilde{Z}^{l}_{\phi j},  \label{decomp}
\end{equation}
where the chiral(antichiral) renormalization superfields
$\tilde{Z}^{l}_{\phi j}(\tilde{\bar Z}^{i}_{\phi k})$ are the wave function
renormalizations and $\Delta^i_j$ is the soft term $(m^2)^i_j$
renormalization.  To find, them expand
$\tilde{Z}^i_j(\tilde{\alpha},\tilde{y},\tilde{\bar y})$ over the
grassmann variables
\begin{equation}
\tilde{Z}^i_j=Z^i_j+D_1Z^i_j \eta+\bar{D}_1Z^i_j
\bar{\eta}+D_2Z^i_j \eta \bar \eta ,
\end{equation}
where  the operator $D_1$  now has to take into account the $\eta$ dependence
of the Yukawa coupling $y^{ijk}$, $\bar{D}_1$ is  conjugated to $D_1$ and
$D_2$ is a second order differential operator which extracts the $\eta\bar
\eta $ dependence of $ \tilde{Z}^i_j.$ They  have the form
\begin{eqnarray}
D_1&=&m_{A_i}\alpha_i\frac{\partial}{\partial \alpha_i}
-A^{ijk}\frac{\partial}{\partial y^{ijk}} , \ \ \
\bar{D}_1=m_{A_i}\alpha_i\frac{\partial}{\partial \alpha_i}
-A_{ijk}\frac{\partial}{\partial y_{ijk}} ,   \label{barD1} \\
D_2&=& \bar{D}_1 D_1 +m_{A_i}^2\alpha_i\frac{\partial }{\partial \alpha_i}
+\frac{1}{2}(m^2)^a_n\left(y^{nbc}\frac{\partial }{\partial y^{abc}} \right.
\label{DD2}\\ && \left. +y^{bnc}\frac{\partial }{\partial y^{bac}}+
y^{bcn}\frac{\partial }{\partial y^{bca}}+
y_{abc}\frac{\partial }{\partial y_{nbc}}+
y_{bac}\frac{\partial }{\partial y_{bnc}}+
y_{bca}\frac{\partial }{\partial y_{bcn}}\right).  \nonumber
\end{eqnarray}
Substituting this into eq.(\ref{decomp}) one has
\begin{eqnarray}
\tilde{Z}^{i}_{\phi j}&=&Z^{i}_{\phi j}+ (Z^{-1}_\phi)^i_k D_1Z^k_j\eta =
Z^{i}_{\phi k}(\delta^k_j+(Z^{-1})^k_l D_1Z^l_j\eta), \\
 \tilde{\bar Z}^{i}_{\phi j}&=&Z^{i}_{\phi j}+ \bar{D}_1Z^i_k
(Z^{-1}_\phi)^k_j \bar \eta = (\delta^i_l+\bar{D}_1Z^i_l(Z^{-1})^l_l \bar
\eta)Z^{l}_{\phi j}, \\
\Delta^i_j&=&(Z^{-1}_\phi)^i_kD_2Z^k_l(Z^{-1}_\phi)^l_j- (Z^{-1}_\phi)^i_k
\bar{D}_1Z^k_l (Z^{-1})^l_n D_1Z^n_m (Z^{-1}_\phi)^m_j, \label{delta}
\end{eqnarray}
where $Z^{i}_{\phi j}$ satisfies
$$Z^{i}_{\phi k}Z^{k}_{\phi j}=Z^i_j$$
and is the square root of $Z^{i}_{ j}$  in the perturbative sense.
The inverse one is
\begin{equation}
(\tilde{Z}^{-1}_\phi)^{i}_{ j}=(\delta^i_k-(Z^{-1})^i_l
D_1Z^l_k\eta)(Z^{-1}_\phi)^{k}_{j} .
\end{equation}

We can now write down the relations between the renormalized and the bare
couplings of a superpotential. One has
\begin{eqnarray}
(\tilde{M}_{Bare})^{ij}&=&\tilde{M}^{kl}(\tilde{Z}^{-1}_\phi)^i_k
(\tilde{Z}^{-1}_\phi)^j_l, \\
(\tilde{y}_{Bare})^{ijk}&=&\tilde{y}^{lmn}(\tilde{Z}^{-1}_\phi)^i_l
(\tilde{Z}^{-1}_\phi)^j_m (\tilde{Z}^{-1}_\phi)^k_n
\end{eqnarray}
or
%\begin{equation}
$$
{M}_{Bare}^{ij}-B_{Bare}^{ij}\eta=(M^{kl}-B^{kl}\eta)(\delta^i_n
-(Z^{-1})^i_m D_1Z^m_n\eta) (Z^{-1}_\phi)^{n}_{k}(\delta^j_p- (Z^{-1})^j_s
D_1Z^s_p\eta) (Z^{-1}_\phi)^{p}_{l}.  $$
%\end{equation}
and analogous for  $y^{ijk}$. Expanding over $\eta$ one has for $M^{ij}$ and
$B^{ij}$
\begin{eqnarray*}
{M}_{Bare}^{ij}&=&M^{kl}(Z^{-1}_\phi)^{i}_{k}(Z^{-1}_\phi)^{j}_{l},
\label{mb}\\
B_{Bare}^{ij}&=&B^{kl}(Z^{-1}_\phi)^{i}_{k}(Z^{-1}_\phi)^{j}_{l} \\
&&+M^{kl}(Z^{-1}_\phi)^i_k (Z^{-1})^j_p D_1Z^p_m (Z^{-1}_\phi)^m_l+
M^{kl}(Z^{-1})^i_m D_1Z^m_n(Z^{-1}_\phi)^n_k (Z^{-1}_\phi)^j_l .
\end{eqnarray*}
The last line can also be rewritten as
\begin{equation}
B_{Bare}^{ij}=B^{kl}(Z^{-1}_\phi)^{i}_{k}(Z^{-1}_\phi)^{j}_{l}+
{M}_{Bare}^{il}(Z^{-1}D_1Z)^j_l+{M}_{Bare}^{li}(Z^{-1}D_1Z)^i_l
\end{equation}
Differentiating  with respect to a scale factor one gets after some
algebra
\begin{eqnarray}
\beta_{M}^{ij} &=&\frac{1}{2}(M^{il}\gamma^j_l+M^{lj}\gamma^i_l), \label{M}
\\ \beta_{B}^{ij} &=& \frac{1}{2}(B^{il}\gamma^j_l+B^{lj}\gamma^i_l)-
(M^{il}D_1\gamma^j_l+M^{lj}D_1\gamma^i_l), \label{bij}
\end{eqnarray}
where $\gamma^i_j$ is the anomalous dimension of the matter chiral field
being the logarithmic derivative of $\ln (Z^{-1})^i_j$.

Analogously,  for the Yukawa coupling $y^{ijk}$ and for the soft triple
coupling $A^{ijk}$ we obtain
\begin{eqnarray}
\beta_{y}^{ijk} &=&\frac{1}{2}(y^{ijl}\gamma^k_l+y^{ilk}\gamma^j_l+
y^{ljk}\gamma^i_l), \label{y} \\
\beta_{A}^{ijk} &=&  \frac{1}{2}(A^{ijl}\gamma^k_l+A^{ilk}\gamma^j_l+
A^{ljk}\gamma^i_l) \label{aijk} \\
&&-(y^{ijl}D_1\gamma^k_l+y^{ilk}D_1\gamma^j_l+y^{ljk}D_1\gamma^i_l).
\nonumber
\end{eqnarray}

Consider now the renormalization of the soft mass term $(m^2)^i_j$. One has
\begin{equation}
(m^2_{Bare})^i_j=(m^2)^i_j-\Delta^i_j
\end{equation}
with $\Delta^i_j$ given by eq.(\ref{delta}). Differentiating it with respect
to a scale factor  and using the explicit form of the second order
differential operator $D_2$,  after some algebra one obtains a simple
expression for the $\beta$ function \begin{equation}
(\beta_{m^2})^i_j=D_2\gamma^i_j \label{mij}
\end{equation}

\subsection{Summary on the soft term renormalizations}

Relations between the rigid and soft terms renormalizations are summarized
in the Table.

\begin{table}[htb]
\begin{center}
\begin{tabular}{|l|l|}
\hline \hline  & \\[-0.2cm]
\hspace*{1.5cm}  The Rigid Terms & \hspace*{1.5cm}  The Soft Terms \\[0.2cm]
\hline  & \\
$\beta_{\alpha_i} =  \alpha_i\gamma_{\alpha_i}$ &
$\beta_{m_{A i}}=D_1\gamma_{\alpha i}$ \\[0.3cm]
$\beta_{M}^{ij} =\frac{1}{2}(M^{il}\gamma^j_l+M^{lj}\gamma^i_l) $&
$\beta_{B}^{ij} = \frac{1}{2}(B^{il}\gamma^j_l+B^{lj}\gamma^i_l)-
(M^{il}D_1\gamma^j_l+M^{lj}D_1\gamma^i_l) $\\[0.3cm]
$\beta_{y}^{ijk} =\frac{1}{2}(y^{ijl}\gamma^k_l+y^{ilk}\gamma^j_l+
y^{ljk}\gamma^i_l)$ &
$\beta_{A}^{ijk} = \frac{1}{2}(A^{ijl}\gamma^k_l+A^{ilk}\gamma^j_l+
A^{ljk}\gamma^i_l)$ \\[0.2cm] &
\hspace*{1cm}
$-(y^{ijl}D_1\gamma^k_l+y^{ilk}D_1\gamma^j_l+y^{ljk}D_1\gamma^i_l)$\\[0.2cm]
&  $(\beta_{m^2})^i_j=D_2\gamma^i_j $ \\[0.3cm]
\hline
\multicolumn{2}{|l|}  {} \\[-0.2cm]
\multicolumn{2}{|l|}
{$D_1= m_{A_i}\alpha_i\frac{\displaystyle \partial}{\displaystyle \partial
\alpha_i} -A^{ijk}\frac{\displaystyle \partial}{\displaystyle \partial
y^{ijk}}\ \ , \hspace{1.7cm}
\bar{D}_1=m_{A_i}\alpha_i\frac{\displaystyle  \partial}{\displaystyle
\partial \alpha_i} -A_{ijk}\frac{\displaystyle \partial}{\displaystyle
\partial y_{ijk}} $ }\\[0.3cm]
\multicolumn{2}{|l|} {$D_2= \bar{D}_1 D_1 +
m_{A_i}^2\alpha_i\frac{\displaystyle \partial }{\displaystyle \partial
\alpha_i} $} \\[0.3cm] \multicolumn{2}{|l|} {\hspace*{0.1cm}
$+\frac{1}{2}(m^2)^a_n\left(y^{nbc}\frac{\displaystyle \partial
}{\displaystyle \partial y^{abc}} +y^{bnc}\frac{\displaystyle \partial
 }{\displaystyle \partial y^{bac}}+ y^{bcn}\frac{\displaystyle \partial
}{\displaystyle \partial y^{bca}}+ y_{abc}\frac{\displaystyle \partial
}{\displaystyle \partial y_{nbc}}+ y_{bac}\frac{\displaystyle \partial
}{\displaystyle \partial y_{bnc}}+ y_{bca}\frac{\displaystyle \partial
}{\displaystyle \partial y_{bcn}}\right)$ }\\[0.3cm] \hline \hline
\end{tabular}
\end{center}
\caption{Relations between the  rigid and soft term renormalizations in a
massless minimal subtraction scheme}
\end{table}

\section{Illustration}

To make the above formulae more clear and to demonstrate how they
work in practice, we consider the renormalization group functions in a
general theory up to two loops. We follow the notation of
ref.~\cite{Jones} except that our $\beta $ functions are half of
those of ref.~\cite{Jones}. Note that all the calculations in
ref.~\cite{Jones} are performed in the framework of dimensional reduction
and the $\overline{MS}$ scheme.

The gauge $\beta $ functions and the anomalous dimensions of matter
superfields in a massless scheme are the functions of dimensionless gauge and
Yukawa couplings of a rigid theory.

\subsection{One-loop renormalization}

In the one-loop order, the renormalization group functions of a rigid
theory are (for simplicity, we consider the case of a single gauge
coupling)
\begin{eqnarray}
\gamma_\alpha^{(1)}&=&\alpha Q, \ \ \
Q=T(R)-3C(G), \\ \gamma^{i\ (1)}_j&=& \frac{1}{2}y^{ikl}y_{jkl}-2\alpha
C(R)^i_j, \end{eqnarray} where the Casimir operators are defined by
$$T(R)\delta_{AB}=Tr(R_AR_B), \ \ C(G)\delta_{AB}=f_{ACD}f_{BCD}, \ \
C(R)^i_j=(R_AR_A)^i_j .$$

Using eqs.(\ref{ai}-\ref{mij}), we construct the renormalization group
functions for the soft terms
\begin{eqnarray}
\beta_{m_A}^{(1)} &=& \alpha m_AQ, \\
\beta_B^{ij\ (1)} &=& \frac{1}{2}B^{il}(\frac{1}{2}y^{jkm}y_{lkm}
-2\alpha C(R)^j_l) \nonumber \\
&&+ M^{il}(\frac{1}{2}A^{jkm}y_{lkm}+2\alpha m_AC(R)^j_l)
+(i\leftrightarrow j), \\
\beta_{A}^{ijk\ (1)}&=&\frac{1}{2}A^{ijl}(\frac{1}{2}y^{kmn}y_{lmn}
-2\alpha C(R)^k_l) \nonumber \\
&&+ y^{ijl}(\frac{1}{2}A^{kmn}y_{lmn}+2\alpha m_AC(R)^k_l) \nonumber \\
&& +(i\leftrightarrow j) +(i\leftrightarrow k), \\
\left[\beta_{m^2}\right]^{i\ (1)}_j &=&
\frac{1}{2}A^{ikl}A_{jkl}-4\alpha m_A^2C(R)^i_j \nonumber \\
&&+\frac{1}{4}y^{nkl}(m^2)^i_ny_{jkl}+\frac{1}{4}y^{ikl}(m^2)^n_jy_{nkl}
+\frac{4}{4}y^{isl}(m^2)^k_sy_{jkl}.
\end{eqnarray}
One can easily see that the resulting formulae coincide with those of
ref.~\cite{Jones} with one exception: we have ignored hereafter all the
tadpole terms assuming that they all are equal to zero because of the
absence of linear terms in the action (see the footnote on p.8).

\subsection{Two-loop renormalization}

In two loops the situation is more tricky. The rigid renormalizations are
\begin{eqnarray}
\gamma_\alpha^{(2)}&=&2\alpha^2C(G)Q-\frac{2\alpha }{r}C(R)^i_j
(\frac{1}{2}y^{jkl}y_{ikl}-2\alpha C(R)^j_i), \ \ r=dim G=\delta_{AA},
\\ \gamma^{i\ (2)}_j&=&-(y^{imp}y_{jmn}+2\alpha C(R)^p_j\delta^i_n)
(\frac{1}{2}y^{nkl}y_{pkl}-2\alpha C(R)^n_p)+2\alpha^2QC(R)^i_j.
\end{eqnarray}
Then, the soft renormalizations are as follows:
\begin{eqnarray}
\beta_{m_A}^{(2)} &=& 4\alpha^2m_AC(G)Q-\frac{2\alpha m_A}{r}C(R)^i_j
(\frac{1}{2}y^{jkl}y_{ikl}-2\alpha C(R)^j_i) \nonumber \\
&&+\frac{2\alpha}{r}C(R)^i_j(\frac{1}{2}A^{jkl}y_{ikl}+2\alpha
m_AC(R)^j_i), \\
\beta_B^{ij\ (2)} &=&
-\frac{1}{2}B^{il}(y^{jkp}y_{lkn}+2\alpha
C(R)^p_l\delta^j_n)(\frac{1}{2}y^{nst}y_{pst}-2\alpha C(R)^n_p)\nonumber \\
&& -M^{il}(A^{jkp}y_{lkn}-2\alpha m_AC(R)^p_l\delta^j_n)
(\frac{1}{2}y^{nst}y_{pst}-2\alpha C(R)^n_p) \nonumber \\
&& -M^{il}(y^{jkp}y_{lkn}+2\alpha C(R)^p_l\delta^j_n)
(\frac{1}{2}A^{nst}y_{pst}+2\alpha m_AC(R)^n_p) \nonumber \\
&&+B^{il}\alpha^2QC(R)^j_l-4M^{il}\alpha^2QC(R)^j_lm_A +
(i\leftrightarrow j), \\
\beta_{A}^{ijk\ (2)} &=&
-\frac{1}{2}A^{ijl}(y^{kmp}y_{lmn}+2\alpha
C(R)^p_l\delta^k_n)(\frac{1}{2}y^{nst}y_{pst}-2\alpha C(R)^n_p)\nonumber \\
&& +A^{ijl}\alpha^2QC(R)^k_l-4y^{ijl}\alpha^2QC(R)^j_lm_A\nonumber \\
&&-y^{ijl}(A^{kmp}y_{lmn}-2\alpha m_AC(R)^p_l\delta^k_n)
(\frac{1}{2}y^{nst}y_{pst}-2\alpha C(R)^n_p) \nonumber \\
&& -y^{ijl}(y^{kmp}y_{lmn}+2\alpha C(R)^p_l\delta^k_n)
(\frac{1}{2}A^{nst}y_{pst}+2\alpha m_AC(R)^n_p) \nonumber \\
&&+ (i\leftrightarrow j)+ (i\leftrightarrow k),\\
\left[\beta_{m^2}\right]^{i\ (2)}_j
&=&-(A^{ikp}A_{jkn}+\frac{1}{2}(m^2)^i_ly^{lkp}y_{jkn}
+\frac{1}{2}y^{ikp}y_{lkn}(m^2)^l_j \nonumber \\
&& +\frac{2}{2}y^{ilp}(m^2)^s_ly_{jsn}+\frac{1}{2}y^{iks}(m^2)^p_sy_{jkn}+
\frac{1}{2}y_{ikp}(m^2)^s_ny_{jks} \nonumber \\
&&+4\alpha m_A^2C(R)^p_j\delta^i_n)(\frac{1}{2}y^{nst}y_{pst}-2\alpha
C(R)^n_p) \nonumber \\
&&-(y^{ikp}y_{jkn}+2\alpha C(R)^p_j\delta^i_n)(\frac{1}{2}A^{nst}A_{pst}
+\frac{1}{4}(m^2)^k_ly^{lst}y_{pst} \nonumber \\
&&+\frac{1}{4}y^{nst}y_{lst}(m^2)^l_p+\frac{4}{4}y^{nlt}(m^2)^s_ly_{pst}
-2\alpha m_A^2C(R)^n_p )\nonumber \\
&&-(A^{ikp}y_{jkn}-2\alpha m_AC(R)^p_j\delta^i_n)
(\frac{1}{2}y^{nst}A_{pst}+2\alpha m_AC(R)^n_p) \nonumber \\
&&-(y^{ikp}A_{jkn}-2\alpha m_AC(R)^p_j\delta^i_n)
(\frac{1}{2}A^{nst}y_{pst}+2\alpha m_AC(R)^n_p) \nonumber \\
&& + 12\alpha^2m_A^2QC(R)^i_j
\end{eqnarray}
Comparing these formulae with those of ref.~\cite{Jones} we find that
they coincide with one exception. The renormalization of $m^2$ in
eq.(16) of ref.~\cite{Jones} has two extra terms, $8g^4SC(R)^i_j$ and
the term proportional to the mass of the so-called
$\varepsilon$-scalars, $\tilde{m}^2$. The first term, though it is not
proportional to $\tilde{m}^2$, still has the same origin. It has
appeared as a result of the $\varepsilon$-scalar mass counterterm in
one loop.

The presence or absence of these terms is renormalization scheme
dependent.  The version of dimensional reduction adopted in
ref.~\cite{Jones} corresponds to diagram by diagram minimal subtraction
of divergences and naturally includes the $\varepsilon$-scalar mass
counterterm. Our substitution being formulated in the superfield
formalism does not contain $\varepsilon$-scalars; no surprise that
the corresponding terms do not appear.

There are two ways to resolve the noticed discrepancy. Either to redefine the
renormalization scheme in such a way that these terms are not present, like
the one in Ref.~\cite{eps-scalar}, or to modify the superfield
renormalization procedure to reproduce these terms
(see also the discusson of this point in Ref. \cite{5authors}).

\subsection{Three-loop soft renormalizations}

To demonstrate the power of the proposed algorithm, we calculate
the three loop gaugino mass renormalization out of a gauge $\beta $
function.  One has in three loops~\cite{Jack}
\begin{eqnarray}
\gamma_\alpha^{(3)}&=&\alpha^3C(G)Q[4C(G)-Q]-\frac{6}{r}
\alpha^3QC(R)^i_jC(R)^j_i \nonumber \\
&+&\frac{3}{r}\alpha^2(y^{ikl}y_{jkl}-4\alpha C(R)^i_j)C(R)^j_sC(R)^s_i
-\frac{2}{r}\alpha^2C(G)(y^{ikl}y_{jkl}-4\alpha C(R)^i_j)C(R)^j_i
\nonumber \\
&+&\frac{3}{2r}\alpha y^{ikm}y_{jkn}(y^{nst}y_{mst}-4\alpha
C(R)^n_m)C(R)^j_i \nonumber \\
&+&\frac{1}{4r}\alpha (y^{ikl}y_{jkl}-4\alpha C(R)^i_j)
(y^{jst}y_{pst}-4\alpha C(R)^j_p)C(R)^p_i .
\end{eqnarray}
The corresponding gaugino mass renormalization is
\begin{eqnarray}
\beta_{m_A}^{(3)} &=& 3\alpha^3C(G)Q[4C(G)-Q]m_A-\frac{18}{r}
\alpha^3QC(R)^i_jC(R)^j_im_A \nonumber \\
&+&\frac{6}{r}\alpha^2(y^{ikl}y_{jkl}-4\alpha C(R)^i_j)C(R)^j_sC(R)^s_im_A
\nonumber \\
&-&\frac{3}{r}\alpha^2(A^{ikl}y_{jkl}+4\alpha C(R)^i_jm_A)
C(R)^j_sC(R)^s_i \nonumber \\
&-&\frac{4}{r}\alpha^2C(G)(y^{ikl}y_{jkl}-4\alpha C(R)^i_j)C(R)^j_i m_A
\nonumber \\
&+&\frac{2}{r}\alpha^2C(G)(A^{ikl}y_{jkl}+4\alpha C(R)^i_jm_A)C(R)^j_i
\nonumber \\
&+&\frac{3}{2r}\alpha y^{ikm}y_{jkn}(y^{nst}y_{mst}-4\alpha
C(R)^n_m)C(R)^j_i m_A \nonumber \\
&-&\frac{3}{2r}\alpha A^{ikm}y_{jkn}(y^{nst}y_{mst}-4\alpha
C(R)^n_m)C(R)^j_i \nonumber \\
&-&\frac{3}{2r}\alpha y^{ikm}y_{jkn}(A^{nst}y_{mst}+4\alpha
C(R)^n_mm_A)C(R)^j_i \nonumber \\
&+&\frac{1}{4r}\alpha (y^{ikl}y_{jkl}-4\alpha C(R)^i_j)
(y^{jst}y_{pst}-4\alpha C(R)^j_p)C(R)^p_i m_A \nonumber \\
&-&\frac{1}{4r}\alpha (A^{ikl}y_{jkl}+4\alpha C(R)^i_jm_A)
(y^{jst}y_{pst}-4\alpha C(R)^j_p)C(R)^p_i  \nonumber \\
&-&\frac{1}{4r}\alpha (y^{ikl}y_{jkl}-4\alpha C(R)^i_j)
(A^{jst}y_{pst}+4\alpha C(R)^j_pm_A)C(R)^p_i.
\end{eqnarray}

\section{Soft Renormalizations in the MSSM}

The general rules described in the previous section can be applied to any
model, in particular to the MSSM. In the case when the field content
and the Yukawa interactions are fixed, it is more useful to deal with
numerical rather than with tensor couplings. Rewriting the
superpotential (\ref{rigid}) and the soft terms (\ref{sofl}) in terms
of group invariants, one has
\begin{equation}
{\cal W}_{SUSY}=\frac{1}{6}\sum_a y_a\lambda^{ijk}_a
\Phi_i\Phi_j\Phi_k+\frac{1}{2}\sum_b M_bh^{ij}_b\Phi_i\Phi_j, \label{pot}
\end{equation}
and
\begin{equation} -{\cal L}_{soft}=\left[\frac 16 \sum_a {\cal A}_a
\lambda^{ijk}_a \phi_i\phi_j\phi_k+ \frac 12 \sum_b {\cal
B}_bh^{ij}_b\phi_i\phi_j + \frac 12 m_{A_j}\lambda_j\lambda_j+h.c.\right]
+(m^2)^j_i\phi^{*i}\phi_j,\label{sof}
\end{equation}
where we have introduced numerical couplings $y_a,M_b,{\cal A}_a$ and
${\cal B}_b$.

Usually, it is assumed that the soft terms obey the universality
hypothesis, i.e. they repeat the structure of a superpotential, namely
\begin{equation}
{\cal A}_a=y_aA_a,\ \ {\cal B}_b=M_bB_b, \ \ (m^2)^i_j=m^2_i\delta^i_j.
\end{equation}
Thus, we have the following set of couplings and soft parameters:
$$
g_j,\ y_a,\ M_b,\ A_a,\ B_b, \ m^2_i,\ m_{A_j}.
$$
Then, the renormalization group $\beta$ functions of a rigid theory
(\ref{ai},\ref{M},\ref{y}) look like (for simplicity, we assume the
diagonal renormalization of matter superfields)
\begin{eqnarray}
\beta_{\alpha_j} &=& \beta_j \equiv \alpha_j\gamma_{\alpha_j}, \\
\beta_{y_a} &=& \frac{1}{2}y_a \sum_iK_{ai}\gamma_i, \\
\beta_{M_b} &=& \frac{1}{2}M_b\sum_iT_{bi}\gamma_i,
\end{eqnarray}
where $\gamma_i$ is the  anomalous dimension of the superfield $\Phi_i$,
$\gamma_{\alpha_j}$ is the anomalous dimension of the gauge superfield
(in some gauges) and numerical matrices $K$ and $T$ specify which
particular fields contribute to a given term in eq.(\ref{pot}).

To get the renormalization of the soft terms, one has to apply the
algorithm of the previous section, eqs.(\ref{ai},\ref{bij},\ref{aijk}).
In terms of numerical couplings it is simplified.

The renormalizations of the soft terms are expressed through those
of a rigid theory in the following way:
\begin{eqnarray}
\beta_{m_{A_j}}&=&D_1\gamma_{\alpha_j}, \label{ma}\\
\beta_{A_a}&=&-D_1\sum_i K_{ai}\gamma_i, \label{A}\\
\beta_{B_b}&=&-D_1\sum_i T_{bi}\gamma_i, \label{B}\\
\beta_{m^2_i}&=&D_2\gamma_i,  \label{m2}
\end{eqnarray}
and the operators $D_1$ and $D_2$ now take the form
\begin{eqnarray}
D_1&=&m_{A_i}\alpha_i \frac{\partial }{\partial \alpha_i}
 -A_aY_a\frac{\partial }{\partial Y_a},  \label{d1}\\
D_2&=&(m_{A_i}\alpha_i \frac{\partial }{\partial \alpha_i}
 -A_aY_a\frac{\partial }{\partial Y_a})^2
+m_{A_i}^2\alpha_i\frac{\partial }{\partial \alpha_i}+
m^2_iK_{ai}Y_a\frac{\partial }{\partial Y_a}. \label{d2}
\end{eqnarray}
where we have used the notation $Y_a\equiv \frac{y_a^2}{16\pi^2}.$

\subsection{Illustration}

To illustrate these rules, we consider as an example one loop
renormalization of the MSSM couplings. Consider for simplicity the
third generation Yukawa couplings only. Then, superpotential is
\begin{equation}
{\cal W}_{MSSM} =(y_tQ^jU^cH_2^i+y_bQ^j{D'}^{c}H_1^i+y_\tau L^jE^cH_1^i
+ \mu H^i_1H^j_2)\epsilon_{ij},
\end{equation}
where $Q,U,D',L$ and $E$ are quark doublet, up-quark, down-quark, lepton
doublet and lepton singlet superfields, respectively, and $H_1$ and
$H_2$ are Higgs doublet superfields. $i$ and $j$ are the $SU(2)$
indices.

The soft terms have a universal form
\begin{eqnarray}
-{\cal L}_{soft-breaking}&=
&\sum_im^2_i|\phi_i|^2+(\frac{1}{2}\sum_a\lambda_a\lambda_a \\
&+& A_ty_tq^ju^ch_2^i+A_by_bq^j{d'}^{c}h_1^i+A_\tau y_\tau l^je^ch_1^i+
B\mu h^i_1h^j_2 + h.c.), \nonumber
\end{eqnarray}
where the small letters denote the scalar components of the
corresponding superfields and $\lambda_a$ are the gauginos. The $SU(2)$
indices are suppressed.

Renormalizations in a rigid theory in the one loop order are given by
the formulae
\begin{eqnarray} \gamma_{\alpha_i}^{(1)}&=&b_i \alpha_i,
\ \ \ i=1,2,3, \ \ \ b_i={\frac{33}{5}, 1,-3}, \\
\gamma_Q^{(1)}&=&Y_t+Y_b-\frac{8}{3}\alpha_3-\frac{3}{2}\alpha_2-
\frac{1}{30}\alpha_1, \\
\gamma_U^{(1)}&=&2Y_t-\frac{8}{3}\alpha_3-\frac{8}{15}\alpha_1, \\
\gamma_D^{(1)}&=&2Y_b-\frac{8}{3}\alpha_3-\frac{2}{15}\alpha_1, \\
\gamma_L^{(1)}&=&Y_\tau -\frac{3}{2}\alpha_2-\frac{3}{10}\alpha_1, \\
\gamma_E^{(1)}&=&2Y_\tau -\frac{6}{5}\alpha_1, \\
\gamma_{H_1}^{(1)}&=&3Y_b+Y_\tau -\frac{3}{2}\alpha_2-\frac{3}{10}\alpha_1,
 \\ \gamma_{H_2}^{(1)}&=&3Y_t -\frac{3}{2}\alpha_2-\frac{3}{10}\alpha_1.
\end{eqnarray}
Consequently, the renormalization group $\beta $ functions are
\begin{eqnarray}
\beta_{\alpha_i}^{(1)}&=&b_i\alpha_i^2, \\
\beta_{Y_t}^{(1)}&=&Y_t(6Y_t+Y_b-\frac{16}{3}\alpha_3-3\alpha_2
-\frac{13}{15}\alpha_1),\\
\beta_{Y_b}^{(1)}&=&Y_b(Y_t+6Y_b+Y_\tau -\frac{16}{3}\alpha_3-3\alpha_2
-\frac{7}{15}\alpha_1),\\
\beta_{Y_\tau }^{(1)} &=&Y_\tau (3Y_b+4Y_\tau -3\alpha_2
-\frac{9}{5}\alpha_1),\\
\beta_{\mu^2}^{(1)}&=&\mu^2(3Y_t+3Y_b+Y_\tau -3\alpha_2-\frac{3}{5}\alpha_1).
\end{eqnarray}
This allows us immediately to write down the soft term renormalizations
\begin{eqnarray}
\beta_{A_t}^{(1)}&=&6Y_tA_t+Y_bA_b+\frac{16}{3}\alpha_3m_{A_3}+
3\alpha_2m_{A_2} +\frac{13}{15}\alpha_1m_{A_1},\\
\beta_{A_b}^{(1)}&=&Y_tA_t+6Y_bA_b+Y_\tau A_\tau
+\frac{16}{3}\alpha_3m_{A_3}+3\alpha_2m_{A_2}+\frac{7}{15}\alpha_1m_{A_1},\\
\beta_{A_\tau }^{(1)} &=&3Y_bA_b+4Y_\tau A_\tau
+3\alpha_2m_{A_2}+\frac{9}{5}\alpha_1m_{A_1},\\
\beta_{B}^{(1)} &=&3Y_tA_t+3Y_bA_b+Y_\tau A_\tau
+3\alpha_2m_{A_2}+\frac{3}{5}\alpha_1m_{A_2}, \\
\beta_{m_{A_i}}^{(1)}&=&\alpha_ib_im_{A_i}, \\
\beta_{m_{Q}^2}^{(1)}&=&Y_t(m_Q^2+m_U^2+m_{H_2}^2+A_t^2)+
Y_b(m_Q^2+m_D^2+m_{H_1}^2+A_b^2)\\
\nonumber && -\frac{16}{3}\alpha_3m_{A_3}^2-
3\alpha_2m_{A_2}^2-\frac{1}{15}\alpha_1m_{A_1}^2, \\
\beta_{m_{U}^2}^{(1)}&=&2Y_t(m_Q^2+m_U^2+m_{H_2}^2+A_t^2)
-\frac{16}{3}\alpha_3m_{A_3}^2-\frac{16}{15}\alpha_1m_{A_1}^2, \\
\beta_{m_{D}^2}^{(1)}&=&2Y_b(m_Q^2+m_D^2+m_{H_1}^2+A_b^2)
-\frac{16}{3}\alpha_3m_{A_3}^2-\frac{4}{15}\alpha_1m_{A_1}^2, \\
\beta_{m_{L}^2}^{(1)}&=&Y_\tau (m_L^2+m_E^2+m_{H_1}^2+A_\tau^2)-
3\alpha_2m_{A_2}^2-\frac{3}{5}\alpha_1m_{A_1}^2, \\
\beta_{m_{E}^2}^{(1)}&=&2Y_\tau (m_L^2+m_E^2+m_{H_1}^2+A_\tau^2)
-\frac{12}{5}\alpha_1m_{A_1}^2,\\
\beta_{m_{H_1}^2}^{(1)}&=&3Y_b(m_Q^2+m_D^2+m_{H_1}^2+A_b^2)+
Y_\tau (m_L^2+m_E^2+m_{H_1}^2+A_\tau^2)\nonumber \\ &&-
3\alpha_2m_{A_2}^2-\frac{3}{5}\alpha_1m_{A_1}^2, \\
\beta_{m_{H_2}^2}^{(1)}&=&3Y_t(m_Q^2+m_U^2+m_{H_2}^2+A_t^2)-
3\alpha_2m_{A_2}^2-\frac{3}{5}\alpha_1m_{A_1}^2,
\end{eqnarray}
which perfectly coincide with those of ref.(\cite{Boer})

\subsection{Two and three-loop gaugino mass renormalization}

We calculate here the two and three loop gaugino mass
renormalization out of a corresponding gauge $\beta $ functions.
The RG $\beta $ functions for the gauge couplings in the MSSM
are~\cite{Ferreira}
\begin{eqnarray}
\beta_{\alpha_i}&=&b_i\alpha_i^2+\alpha_i^2\left(\sum_jb_{ij}\alpha_j-\sum_f
a_{if}Y_f \right) \nonumber \\
&& +\alpha_i^2\left[\sum_{jk}b_{ijk}\alpha_j\alpha_k
-\sum_{jf}a_{ijf}\alpha_jY_f+\sum_{fg}a_{ifg}Y_fY_g \right] + ... ,
\end{eqnarray}
where $Y_f$ means $Y_t,Y_b$ and $Y_\tau $ and the coefficients
$b_i,b_{ij},a_{if},b_{ijk},a_{ijf}$ and $a_{ifg}$ are given in
ref.~\cite{Ferreira}.

For the gaugino masses we have
\begin{eqnarray}
\beta_{m_{A_i}}&=&b_i\alpha_im_{A_i}+
\alpha_i\left(\sum_jb_{ij}\alpha_j(m_{A_i}+m_{A_j})-\sum_f
a_{if}Y_f(m_{A_i}-A_f) \right) \nonumber \\
&&+\alpha_i\left[\sum_{jk}b_{ijk}\alpha_j\alpha_k(m_{A_i}+m_{A_j}+m_{A_k})
-\sum_{jf}a_{ijf}\alpha_jY_f(m_{A_i}+m_{A_j}-A_f)\right. \nonumber \\
&& \left. +\sum_{fg}a_{ifg}Y_fY_g (m_{A_i}-A_f-A_g) \right] + ... ,
\end{eqnarray}

The same formulae can easily be obtained for the other soft terms. We
do not write them down here due to the lack of space. Instead, the
explicit formulae in the case when only top Yukawa coupling and
$\alpha_3$ are retained are presented in Appendix.

\section*{Appendix. Three-loop renormalizations in the MSSM}

In this section, we present explicit formulae for rigid and soft term
renormalizations in the MSSM in  the three-loop approximation in the
case when we retain only $\alpha_3$ and top Yukawa coupling $Y_t$.

The rigid renormalizations are~\cite{Ferreira}
\begin{eqnarray}
\beta_{\alpha_3}&=&-3\alpha_3^2+\alpha_3^2(14\alpha_3-4Y_t)+
\alpha_3^2[\frac{347}{3}\alpha_3^2-\frac{104}{3}\alpha_3Y_t+30Y_t^2], \\
\gamma_t&=&(2Y_t-\frac{8}{3}\alpha_3)-(8Y_t^2+\frac{8}{9}\alpha_3^2)
+[(30+12\zeta_3)Y_t^3 \\ &+& (\frac{16}{3}+96\zeta_3)Y_t^2\alpha_3-
(\frac{64}{3}+\frac{544}{3}\zeta_3)Y_t\alpha_3^2+(\frac{2720}{27}+320
\zeta_3)\alpha_3^3],\nonumber \\
\gamma_b&=&-\frac{8}{3}\alpha_3-\frac{8}{9}\alpha_3^2
+ [-\frac{80}{3}Y_t\alpha_3^2+(\frac{2720}{27}+320
\zeta_3)\alpha_3^3], \\
\gamma_Q&=&(Y_t-\frac{8}{3}\alpha_3)-(5Y_t^2+\frac{8}{9}\alpha_3^2)+
[(15+6\zeta_3)Y_t^3 \\ &+&(\frac{40}{3}+48\zeta_3)Y_t^2\alpha_3-
(\frac{72}{3}+\frac{272}{3}\zeta_3)Y_t\alpha_3^2+(\frac{2720}{27}+320
\zeta_3)\alpha_3^3],\nonumber \\
\gamma_{H_2}&=&(3Y_t)-(9Y_t^2-16Y_t\alpha_3) \\ &+&
[(57+18\zeta_3)Y_t^3+(72-144\zeta_3)Y_t^2\alpha_3-
(\frac{160}{3}+16\zeta_3)Y_t\alpha_3^2], \nonumber\\
\beta_{Y_t}&=&Y_t\left\{(6Y_t-\frac{16}{3}\alpha_3)-(22Y_t^2-16Y_t\alpha_3
+\frac{16}{9}\alpha_3^2)\right. + [(102+36\zeta_3)Y_t^3 \\ &+& \left.
\frac{272}{3}Y_t^2\alpha_3-
(\frac{296}{3}+288\zeta_3)Y_t\alpha_3^2+(\frac{5440}{27}+640
\zeta_3)\alpha_3^3]\right\}, \nonumber\\
\beta_{\mu^2}&=&\mu^2\left\{3Y_t-(9Y_t^2-16Y_t\alpha_3)\right.  \\ &+&
\left. [(57+18\zeta_3)Y_t^3+(72-144\zeta_3)Y_t^2\alpha_3-
(\frac{160}{3}+16\zeta_3)Y_t\alpha_3^2]\right\} \nonumber,
\end{eqnarray}

The corresponding soft term renormalizations read
\begin{eqnarray}
\beta_{M_3}&=&-3\alpha_3M_3+28\alpha_3^2M_3-4Y_t\alpha_3(M_3-A_t)\\ &+&
347\alpha_3^3M_3-\frac{104}{3}\alpha_3^2Y_t(2M_3-A_t)
+30\alpha_3Y_t^2(M_3-2A_t),\nonumber\\
\beta_{A_t}&=&(6Y_tA_t+\frac{16}{3}\alpha_3M_3)-[44Y_t^2A_t-16Y_t\alpha_3
(A_t-M_3) -\frac{32}{9}\alpha_3^2M_3] \\ &+&
[(306+108\zeta_3)Y_t^3A_t+\frac{272}{3}Y_t^2\alpha_3(2A_t-M_3)\nonumber \\
&-&(\frac{296}{3}+288\zeta_3)Y_t\alpha_3^2(A_t-2M_3)
-3(\frac{5440}{27}+640 \zeta_3)\alpha_3^3M_3],\nonumber \\
\beta_{B}&=&3Y_tA_t-[18Y_t^2A_t-16Y_t\alpha_3(A_t-M_3)] \\ &+&
[(171+54\zeta_3)Y_t^3A_t+(72-144\zeta_3)Y_t^2\alpha_3(2A_t-M_3)\nonumber \\
& -& (\frac{160}{3}+16\zeta_3)Y_t\alpha_3^2(A_t-2M_3)],\nonumber \\
\beta_{m^2_t}&=&2Y_t(m^2_t+m^2_Q+m^2_{H_2}+A_t^2)-\frac{16}{3}\alpha_3M_3^2
-16Y_t^2(m^2_t+m^2_Q+m^2_{H_2}+2A_t^2)\nonumber \\
&-& \frac{16}{3}\alpha_3^2M_3^2+
3(30+12\zeta_3)Y_t^3(m^2_t+m^2_Q+m^2_{H_2}+3A_t^2)
\nonumber \\&+&(\frac{16}{3}+96\zeta_3)Y_t^2\alpha_3
[(2A_t-M_3)^2+2(m^2_t+m^2_Q+m^2_{H_2})+M_3^2] \nonumber \\ &-&
(\frac{64}{3}+\frac{544}{3}\zeta_3)Y_t\alpha_3^2
[(A_t-2M_3)^2+(m^2_t+m^2_Q+m^2_{H_2})+2M_3^2]  \nonumber \\
&+&12(\frac{2720}{27}+320\zeta_3)\alpha_3^3M_3^2, \\
\beta_{m^2_b}&=&-\frac{16}{3}\alpha_3M_3^2-\frac{16}{3}\alpha_3^2M_3^2
\nonumber \\ &-& \frac{80}{3}Y_t\alpha_3^2
[(A_t-2M_3)^2+(m^2_t+m^2_Q+m^2_{H_2})+2M_3^2] \nonumber \\
&+& 12(\frac{2720}{27}+320\zeta_3)\alpha_3^3M_3^2,  \\
\beta_{m^2_Q}&=&Y_t(m^2_t+m^2_Q+m^2_{H_2}+A_t^2)-\frac{16}{3}\alpha_3M_3^2
-10Y_t^2(m^2_t+m^2_Q+m^2_{H_2}+2A_t^2)\nonumber \\ &-&
\frac{16}{3}\alpha_3^2M_3^2+
3(15+6\zeta_3)Y_t^3(m^2_t+m^2_Q+m^2_{H_2}+3A_t^2) \nonumber \\
&+&(\frac{40}{3}+48\zeta_3)Y_t^2\alpha_3
[(2A_t-M_3)^2+ 2(m^2_t+m^2_Q+m^2_{H_2})+M_3^2] \nonumber \\
&-& (\frac{72}{3}+\frac{272}{3}\zeta_3)Y_t\alpha_3^2
[(A_t-2M_3)^2+(m^2_t+m^2_Q+m^2_{H_2})+2M_3^2] \nonumber \\
&+&12(\frac{2720}{27}+320\zeta_3)\alpha_3^3M_3^2, \\
\beta_{m^2_{H_2}}&=&3Y_t(m^2_t+m^2_Q+m^2_{H_2}+A_t^2)
-18Y_t^2(m^2_t+m^2_Q+m^2_{H_2}+2A_t^2) \nonumber \\ &+& 16Y_t\alpha_3
[(A_t-M_3)^2+(m^2_t+m^2_Q+m^2_{H_2})+M_3^2]\nonumber \\ &+&
3(57+18\zeta_3)Y_t^3(m^2_t+m^2_Q+m^2_{H_2}+3A_t^2) \nonumber \\
&+&(72-144\zeta_3)Y_t^2\alpha_3
[(2A_t-M_3)^2+ 2(m^2_t+m^2_Q+m^2_{H_2})+M_3^2] \nonumber \\ &-&
(\frac{160}{3}+16\zeta_3)Y_t\alpha_3^2
[(A_t-2M_3)^2+(m^2_t+m^2_Q+m^2_{H_2})+ 2M_3^2],
\end{eqnarray}
where $M_3\equiv m_{A_3}$.

\vspace{1cm}

{\large \bf Acknowledgements}

\vspace{0.3cm}

D.K. is grateful to D.Maison and K.Chetyrkin for valuable discussions.
Financial support from RFBR grant \# 96-02-17379a and RFBR-DFG
grant \# 00082G and DFG grant \# 436 RUS 113/335/O(R)  is kindly
acknowledged. D.K. and I.K. would like to express their gratitude to
the University of Karlsruhe where part of this work has been done.

\vspace{1cm}

\noindent {\em \bf Note added} \\
When this paper has already been finished we became aware of the
paper \cite{NewJ}, where similar results were obtained.
Our results coincide with their ones in many points. Except for
the relation between the gauge coupling $\beta-$function and the
gaugino $\beta-$function, the authors of Ref. \cite{NewJ}
have deduced the same relations between the renormalization
group functions of the soft and rigid theories through the
differential operators $D_1$ and $D_2$ starting from the
Yamada's rules \cite{Yamada}. As for the $\beta_{\alpha}$--$\beta_{m_A}$
relation, their starting point was the Hisano--Shifman formula
\cite{Shifman}. Our approach is based on a consideration of the
soft theory as a rigid one embedded into the
external $x-$independent superfields, that are the charges and masses
of the theory. The Yamada's rules together with
the operator constructions $D_1$ and $D_2$ are just the
technical consequences of this approach.

\end{document}